\documentclass[reprint,prl,twocolumn,showpacs,superscriptaddress,aps,longbibliography]{revtex4-1}

\usepackage[utf8]{inputenc}

\usepackage[colorlinks=true,citecolor=blue]{hyperref} 
\usepackage{graphicx}
\usepackage{bm}
\usepackage{amsmath}
\usepackage{amssymb}

\usepackage{graphicx}
\DeclareGraphicsExtensions{.png,.jpg,.eps}

\usepackage{float}
\usepackage{xcolor}
\usepackage{amsmath}
\usepackage[printwatermark]{xwatermark}
\usepackage{float}
\usepackage[colorlinks=true,citecolor=blue]{hyperref}
\DeclareMathAlphabet\mathbfcal{OMS}{cmsy}{b}{n}

\usepackage{xcolor}

\begin{document}

\title{Spectroscopy of the frustrated quantum antiferromagnet Cs$_2$CuCl$_4$}

\author{Adolfo O. Fumega}
\affiliation{Departamento de Física Aplicada, Universidade de Santiago de Compostela, E-15782 Campus Sur s/n, Santiago de Compostela, Spain}
\affiliation{Instituto de Investigacións Tecnolóxicas, Universidade de Santiago de Compostela, E-15782 Campus Sur s/n, Santiago de Compostela, Spain}

\author{D. Wong}
\affiliation{\mbox{Helmholtz-Zentrum Berlin f\"ur Materialien und Energie, Albert-Einstein-Stra\ss\hspace{0pt}e 15, 12489 Berlin, Germany}}

\author{C. Schulz}
\affiliation{\mbox{Helmholtz-Zentrum Berlin f\"ur Materialien und Energie, Albert-Einstein-Stra\ss\hspace{0pt}e 15, 12489 Berlin, Germany}}

\author{F. Rodríguez}
\affiliation{MALTA TEAM, DCITIMAC, Facultad de Ciencias, Universidad de Cantabria, 39005 Santander, Spain}

\author{S. Blanco-Canosa}
\email{sblanco@dipc.org}
\affiliation{Donostia International Physics Center (DIPC), San Sebastián, Spain}
\affiliation{IKERBASQUE, Basque Foundation for Science, 48013 Bilbao, Spain}

\begin{abstract}

We investigate the electronic structure of Cs$_2$CuCl$_4$, a material discussed in the framework of a frustrated quantum antiferromagnet, 
by means of resonant inelastic x-ray scattering (RIXS) and Density Functional Theory (DFT). From the non-dispersive highly localized \textit{dd} excitations, 
we resolve the crystal field splitting of the Cu$^{2+}$ ions in a strongly distorted tetrahedral coordination. This allows us to model the RIXS spectrum within the 
Crystal Field Theory (CFT), assign the \textit{dd} orbital excitations and retrieve experimentally the values of the crystal field splitting parameters \textit{D}$_q$, 
\textit{D}$_s$ and \textit{D}$_{\tau}$. The electronic structure obtained \textit{ab-initio} agrees with the RIXS spectrum and modelled by CFT, 
highlighting the potential of combined spectroscopic, cluster and DFT calculations to determine the electronic ground state of complex materials.      

\end{abstract}

\maketitle

\section{Introduction}
Geometrically frustrated spin textures are excellent candidates to search for new forms of exotic magnetism, like quantum spin liquids \cite{Sav16} and topological magnetic ordering \cite{Fert17}. The magnetic order and its excitations, resulting from the geometrically frustrated triangular, Kagomé or honeycomb lattices are the focus of current and intense research \cite{Tak19}. This rich phenomenology is also a manifestation of the interplay between multiple energy scales - magnetic exchange, Coulomb interaction, orbital excitations and crystal field splitting, $\Delta^{\text{CF}}$, as has been observed in the cubic vanadates \cite{Khal01} and superconducting cuprates \cite{Kei15}. 

\begin{figure*}
\begin{center}
\includegraphics[width=1.9\columnwidth,draft=false]{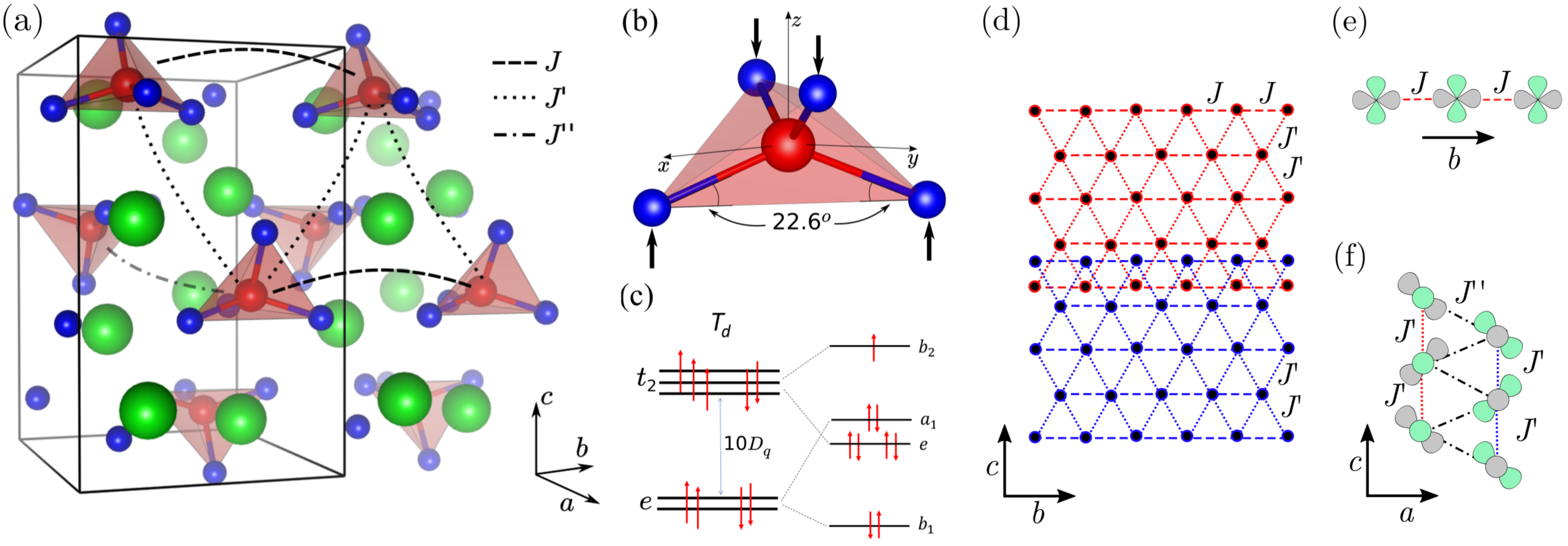}
\caption{(a) Orthorhombic unit cell of Cs$_2$CuCl$_4$. Green, red and blue balls denote Cs, Cu and Cl atoms, respectively. The tetrahedral coordination of Cu is highlighted. The intrachain \textit{J} (dashed), the interchain \textit{J}' (doted) and the interplane \textit{J}'' (dot-dashed) magnetic exchange interactions have been depicted. Notice the triangular layers formed by the \textit{J} and \textit{J}' interactions. These are schematically shown in panel (d). Notice also the quasi-1D Heisenberg chains along the \textit{b} axis formed by the \textit{J} interactions. This is better seen in panel (e). (b) Locally distorted tetrahedral environment of Cu atom. Black arrows along the $z$-direction represent the distortion of the tetrahedron to a square planar situation. The Cl-Cu-Cl angles have been represented to denote the highly distorted scenario. (c) Crystal field splitting of the Cu$^{2+}$ ions in the distorted tetrahedral environment, where the \textit{e} orbitals lower their energy with respect to the \textit{t}$_{2}$. The Jahn-Teller distortion further splits the energy levels with \textit{b}$_{1}$ (\textit{d}$_{x^2-y^2}$), \textit{e} (\textit{d}$_{xz}$/\textit{d}$_{yz}$), \textit{a}$_{1}$ (\textit{d}$_{z^2}$) and \textit{b}$_{2}$ (\textit{d}$_{xy}$) symmetries. Their relation to the crystal field parameters \textit{D}$_q$, \textit{D}$_s$ and \textit{D}$_{\tau}$ is provided by eq. (\ref{onsite_ener}).
(d) Representation of two planes, red and blue, each of them defined by the $J$ and $J'$ exchange constants. The sites were the magnetic moments lay are represented as black dots. (e) Schematic of the 1D-Heisenberg chain along the \textit{b} axis defined by the $J$ couplings. A direct overlap between the \textit{b}$_{2}$ (\textit{d}$_{xy}$) orbitals (depicted in green and grey) can be seen in this direction. (f) Schematic of the $J'$ and $J''$ couplings. Note how the \textit{b}$_{2}$ orbital coupled by $J'$ are tilted 90$^{\circ}$ from one chain to the other along the \textit{c} axis. A more complex interplane geometry can be seen along the \textit{a} axis.
}
\label{Fig1}
\end{center}
\end{figure*}

Among the solids showing frustrated magnetism Cs$_2$CuCl$_4$ has attracted great attention. Below the Néel temperature T$_N$= 0.61 K, Cs$_2$CuCl$_4$ shows a long-range 
antiferromagnetic (AF) order with an incommensurate spiral wave vector. Between T$_N$ and the Curie temperature $T_C$=2.65 K, a 
spin-liquid phase with strong in-chain spin correlations has been discussed by means of specific heat \cite{Tok06} and electron spin resonance (ESR) \cite{Zvy14}. 
Neutron scattering \cite{Col01} experiments have uncovered an extensive two-spinon continuum and revealed a weaker interchain exchange coupling, \textit{J'}/\textit{k}$_B$= 1.4 K, 
between Cu atoms than the intrachain interaction, \textit{J}/\textit{k}$_B$=4.7 K, along \textit{b} axis \cite{Col03}, as depicted in Fig. \ref{Fig1}(a). 
In addition, interlayer coupling in Cs$_2$CuCl$_4$ is smaller than \textit{J} and \textit{J'} by more than one order of magnitude, \textit{J''}= 0.13 K, in good agreement
with the theoretical description of a quasi-1D weakly coupled \textit{S}=1/2 Heisenberg chain \cite{Bal10}.

A quantitative description of spin exchange interactions in frustrated magnets, as in Cs$_2$CuCl$_4$, requires a precise knowledge of the crystal-field ground state ($\Delta^{\text{CF}}$) and electron-orbital excitations. 
However, the determination of the orbital ground state, the value of $\Delta^{\text{CF}}$ and the Jahn-Teller distortion ($\Delta^{\text{JT}}$) of the tetrahedrally coordinated Cu$^{2+}$ ions in Cs$_2$CuCl$_4$ are still missing and their relationship with the magnetic anisotropy ( \textit{J}$\approx 3\times$\textit{J}') needs further study.
Experimentally, inelastic neutron scattering \cite{Xiao13,Bab15} and x-ray absorption spectroscopy (XAS) \cite{But16} have been used to study the magnitude of the crystal field splitting and the \textit{dd} orbital excitations, but the information is sometimes hampered by phonons, small amount of crystals or by the metal-ligand hybridization. Theoretically, \textit{ab initio} density functional theory (DFT) methods provide only a single-electron approximation of excited states, often in disagreement with experimental spectroscopic measurements. This is further complicated by strong quantum fluctuations, associated with spin-1/2 degrees of freedom in the highly localized regime and geometrical frustration that can give rise to entangled quantum phases. 
Interestingly, changes in the symmetry of the occupied orbitals and large orbiton dispersions have been observed in several cuprates \cite{Schla12,Bis15} and titanates \cite{Ulr09} in the energy range between 0.5-2 eV by means of resonant inelastic x-ray scattering (RIXS). On the other side, the study of orbital excitations in non-oxide  materials is rather scarce and the consequences for the physics of \textit{quasi} one dimensional systems are less explored.

Here, we have carried out a comprehensive spectroscopic and theoretical study of the quantum antiferromagnet Cs$_2$CuCl$_4$ by means of RIXS, DFT and cluster calculations. We have been able to resolve the crystal field splitting, and show that the \textit{dd} orbital excitations are well modeled assuming a strongly distorted tetrahedral symmetry with crystal field parameters \textit{D}$_q$= -0.120 eV, \textit{D}$_s$= 0.085 eV and \textit{D}$_{\tau}$= -0.165 eV. Moreover, despite the strong hybridization between the Cu$^{2+}$ and Cl$^-$ ions, we find that the RIXS spectra describes a strongly ionic character of Cu-Cl bond, which may explain the absence of dispersion of the \textit{dd} excitations. On a broader picture, our results exemplify how RIXS, DFT and cluster calculations can be combined to study the ground state of complex materials.

\section{Experimental and Computational Methods}
Single crystals of Cs$_2$CuCl$_4$ were grown by slow evaporation at 300 K from an acidic (HCl) solution containing a 2:1 stoichiometric ratio of the CsCl and CuCl$_2$.H$_2$O. The quality of the single crystals was checked by x-ray diffraction, resulting in lattices parameters \textit{a}= 9.77 \AA, \textit{b}= 7.61 \AA, \textit{c} = 12.41 \AA, and Raman scattering \cite{Jara19}. DFT calculations \cite{dft,dft_2} were performed using the all-electron, full-potential {\sc wien2k} code \cite{wien} based on the augmented plane wave plus local orbital (APW+lo) basis set \cite{sjo}.
The generalized gradient approximation (GGA) in the Perdew-Burke-Ernzerhof \cite{PBE} scheme was used for the exchange correlation functional, with a fully converged \textit{k}-mesh of R$_{mt}$K$_{max}$=7.0 and muffin-tin radii of 2.5, 2.22, 1.91 a.u. for Cs, Cu and Cl, respectively. 
RIXS experiments were performed at the U41-PEAXIS beamline at BESSY II at 20 K and combined energy resolution, $\Delta$\textit{E} $\approx$ 150 meV. To minimize the elastic background, the measurements were carried out with horizontal polarization and the scattering angle 2$\theta$ was set to 110$^{\circ}$. The elastic (zero-energy loss) peak position was determined from the elastic scattering spectra of carbon tape. The Cs$_2$CuCl$_4$ crystal was plate-like with a flat surface normal to the (010) direction. 

\section{Results and Discussion}
The crystal structure of Cs$_2$CuCl$_4$ is orthorhombic with space group \textit{Pnma}. Each Cu$^{2+}$ atom is surrounded by 4 Cl$^-$ ions in a distorted tetrahedral coordination (Fig. 1(a)), with Cu-Cl bond length of 2.25\AA, and Cl-Cu-Cl angle of 22.6$^o$, smaller than the perfect tetrahedral value of 45$^o$, Fig. \ref{Fig1}(b), due to the large distortion of the tetrahedron towards a square planar scenario.
As shown in Fig. \ref{Fig1}(c), the tetrahedral environment splits the \textit{d}-levels into low energy 2-fold \textit{e} and high energy 3-fold degenerate \textit{t}$_{2}$ orbitals (Fig. 1(c)), which are further Jahn-Teller distorted towards a lower symmetry, \textit{D}$_{2d}$ point group \cite{Lam86}. This chemical environment provides a spatially anisotropic spin-1/2 triangular antiferromagnet of Cu-Cl chains on a geometrical \textit{bc} plane, bonded  by Cl$^-$ ions along the \textit{a} direction. Magnetic susceptibility, showing no trace of magnetic order down to 5 K (not shown), is in good agreement with a weakly frustrated magnet \cite{Wei05}.

\begin{figure}[h!]
\begin{center}
\includegraphics[width=\columnwidth,draft=false]{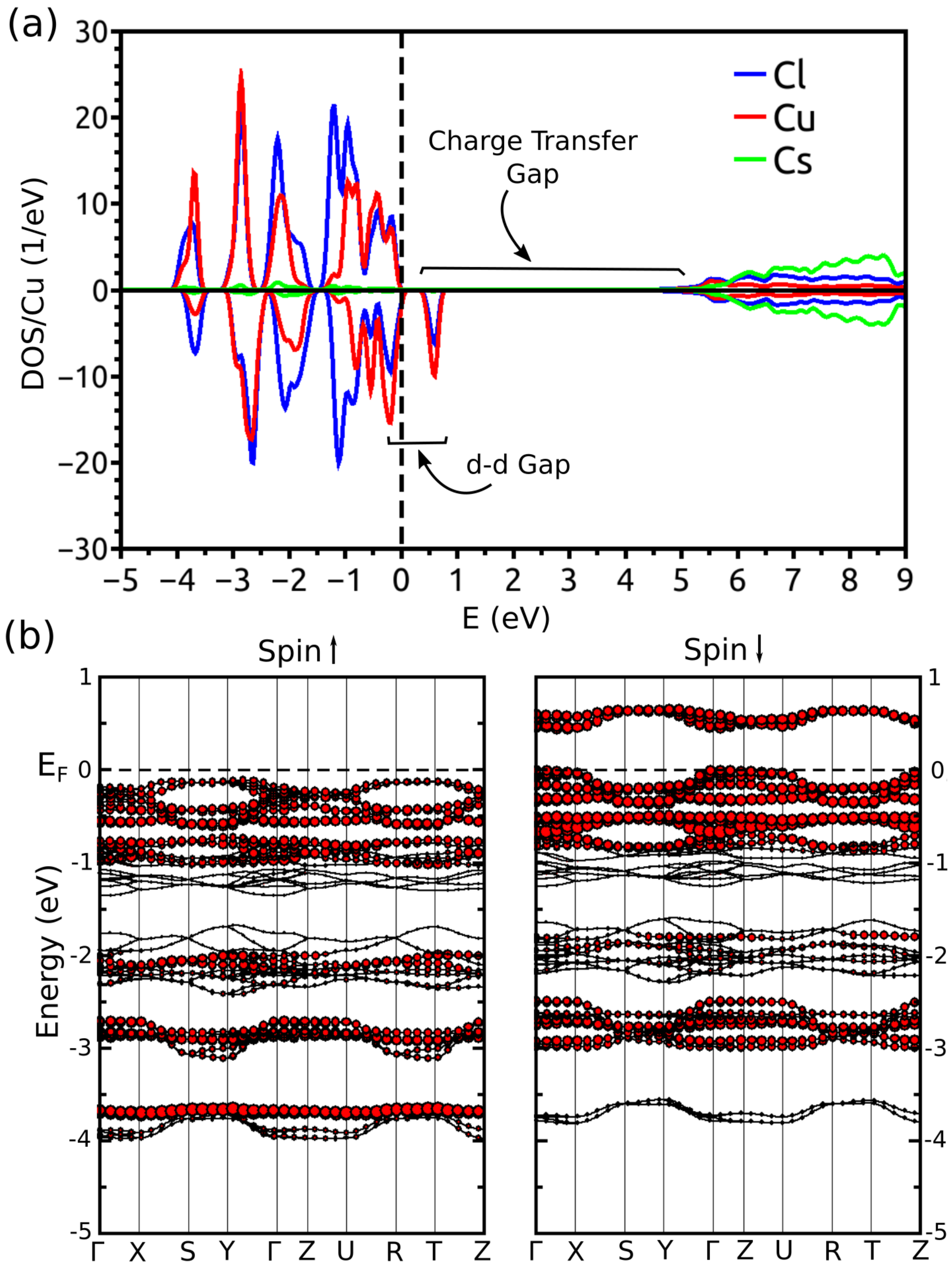}
\caption{Summary of the DFT calculations (a) Projected density of states (DOS) for the ferromagnetic structure of Cs$_2$CuCl$_4$, Cs (Cu, Cl) atom in green (red, blue). Both charge transfer and \textit{dd} energy gaps have been highlighted. (b) Energy bands for the majority (minority) spin channel on the left (right) panel. The \textit{d}-orbital character of the Cu atom is depicted as red circles.  
}
\label{Fig2}
\end{center}
\end{figure}

Figure \ref{Fig2} summarizes the DFT results of the electronic structure of Cs$_2$CuCl$_4$ for the simple unit cell depicted on Fig. \ref{Fig1}(a). This structure leads to a ferromagnetic state, since there is only one non-equivalent Cu atom. 
Calculations assuming a non-magnetic ground state severely underestimate the charge gap and directly give a metallic solution. We have found that calculations on a $1\times 2\times 1 $ supercell provide an antiferromagnetic order as the lowest energy solution for the ground state, in agreement with the previous report of Foyevtsova et al. \cite{Val11}. 
However, both magnetic orders lead to the same electronic energy levels, and consequently to the same energy gaps. Therefore, this guarantees that the ferromagnetic solution can be used to analyze the electronic spectrum of Cs$_2$CuCl$_4$. The usage of the ferromagnetic state entails several advantages for the analysis of the spectrum. First, it makes the calculation computationally affordable, especially for the wannierization that will be carried out. A reduction of the Bloch subspace from 136 to 68 bands is achieved when considering the ferromagnetic state instead of the antiferromagnetic one. Apart from that, the ferromagnetic solution allows for a better visualization of the Cu spin state, due to the asymmetric spin channel band occupation. This allows a direct comparison between the crystal field splitting (Fig. \ref{Fig1}c), the DFT band structure (Fig. \ref{Fig2}b) and the RIXS spectrum (Fig. \ref{Fig4}b).


Between -4 and 0 eV (Fig. \ref{Fig2}(a)) the density of states (DOS) shows a strong Cu 3\textit{d} and Cl 3\textit{p} character, while the Cs bands do not contribute at the Fermi level and, therefore, do not hybridize with the CuCl$_4$ cluster. The Cu and Cl bands are separated from the next unoccupied states by a gap of 4.5 eV, having mostly Cs character. We call this charge transfer gap and corresponds to the energy difference between the CuCl$_4$ cluster and the Cs ions. On the other hand, the \textit{dd} gap appears below 1 eV and represents the energy gap between the Fermi level and the empty \textit{d}$_{xy}$ orbital. Figure 2(b) delves into the band structure of Cs$_2$CuCl$_4$. Bonding and antibonding bands show up between -4 and -3 eV and -1 and 1 eV, respectively.
Moreover, our calculations predict  a spin $1/2$ for the Cu atoms resulting from the asymmetric spin channel occupation. 
It can be clearly seen in Fig. \ref{Fig2}b that the minority spin channel undergoes a shift towards higher energies. Whether this shift is a consequence of the Coulomb repulsion or the exchange interactions cannot be determined from our analysis. However, considering that the the exchange interaction is one order of magnitude smaller than Coulomb repulsion, this could be the dominant term in the substantial energy shift we found in our calculations.

\begin{figure*}
\begin{center}
\includegraphics[width=1.9\columnwidth,draft=false]{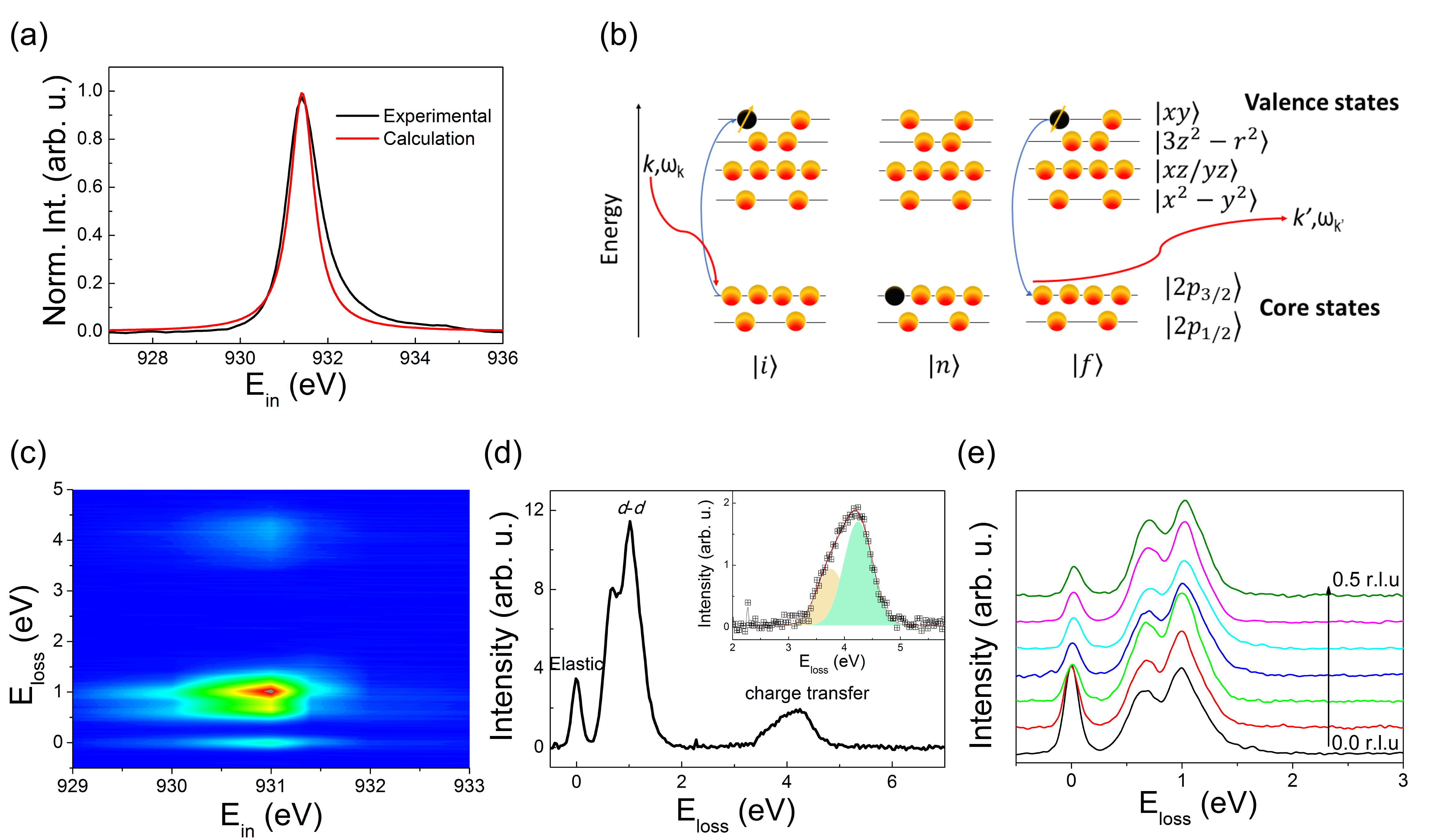}
\caption{(a) Experimental (black) and calculated (red) x-ray absorption (XAS) of Cs$_2$CuCl$_4$. (b) Schematics of the RIXS process showing the initial, intermediate and final steps. (c) RIXS map plotting the energy loss, E$_\text{loss}$, against the incoming energy, E$_\text{in}$, for Cs$_2$CuCl$_4$ at 20 K. The maximum intensity of the inelastic features appears at E$_\text{in}$= 931 eV. (d) Close up view of the RIXS scan at E$_\text{in}$=931 eV, highlighting the elastic, \textit{dd} excitations and the charge gap in the inset. (e) In-plane momentum dependence of the \textit{dd} excitations, showing no orbital dispersion.
}
\label{Fig3}
\end{center}
\end{figure*}

Having carried out a theoretical description of the electronic structure of Cs$_2$CuCl$_4$, we proceed with the analysis of the RIXS spectra. The fast improvement of the RIXS instrumentation in energy resolution of soft x-ray RIXS has allowed the study of the low energy electronic properties of correlated systems, giving detailed information of collective magnetic, charge and orbital excitations \cite{More11}. Figure 3(b) illustrates the case of the Cu$^{2+}$ ion with a  3\textit{d}$^9$ electronic configuration in a tetrahedral crystal field (\textit{T}$_d$), with a hole in a $\vert$\textit{xy}$\rangle$ state.
In the initial step of RIXS, an x-ray photon excites a 2\textit{p}$_{3/2}$ electron from the ground state, $\vert$\textit{i}$\rangle$, into the 3\textit{d} shell (intermediate state, $\vert$\textit{n}$\rangle$) filling the $\vert$\textit{xy}$\rangle$ orbital and creating an excited core-hole state. This process corresponds to the x-ray absorption, XAS, at 931.5 eV (black curve in figure 3a).
In the final step ($\vert$\textit{f}$\rangle$), the core hole is annihilated via decay towards the ground state (elastic scattering, E$_\text{loss}$=0 eV) or an excited state (magnons, phonons, \textit{dd} transitions) \cite{Ament11}. 

\begin{figure}
\begin{center}
\includegraphics[width=1.0\columnwidth,draft=false]{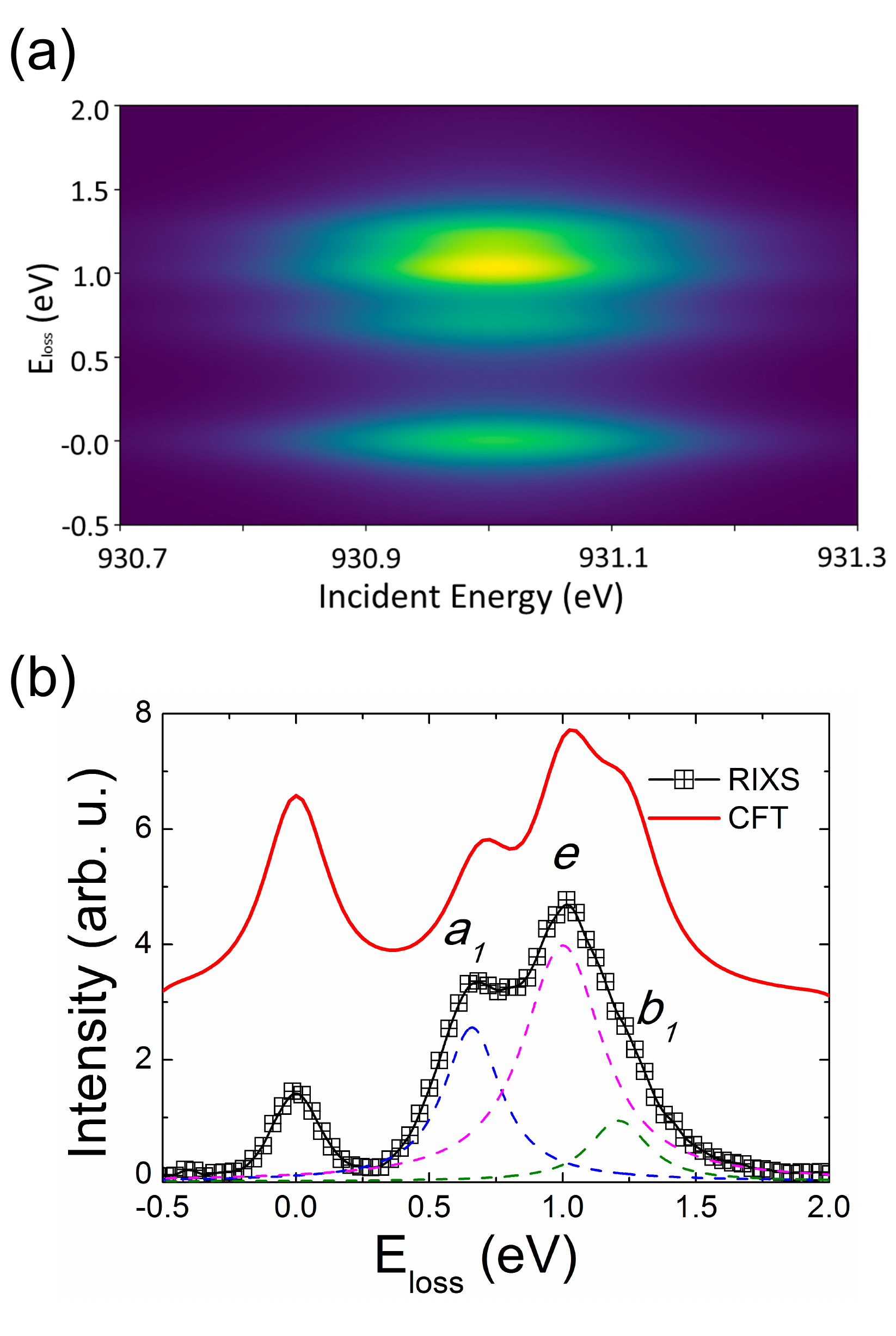}
\caption{(a) Calculated RIXS intensity map within the CFT theory, highlighting the crystal field excitations. (b) Comparison of the experimental and theoretical RIXS spectrum for 931 eV incident energy. Labeled are the \textit{dd} transitions. The calculated RIXS spectrum is vertically shifted for clarity.
}
\label{Fig4}
\end{center}
\end{figure}

Figure \ref{Fig3}(c) displays the incident energy (E$_\text{in}$) vs energy loss (E$_\text{loss}$) RIXS map. The maximum intensity of the resonant features is observed at E$_\text{in}$= 931 eV (Fig. 3(d)), 0.5 eV below the maximum of the \textit{L}$_3$ absorption edge (Figure 3(a)). In nice agreement with the DFT calculations, the charge transfer excitations resulting from the CuCl$_4$ cluster to the Cs 6\textit{s} states are observed as a broad band at E$_\text{loss}$= 4 eV, matching the optical gap observed by absorption spectroscopy \cite{Rod98}. As shown in the inset of Fig. 3(d), this charge transfer gap displays 2 broad bands at 3.7 and 4.2 eV, corresponding to transitions from the Fermi level and the \textit{d}$_{xy}$ orbital above the Fermi level (Fig. 2(b)) to the upper Cs 6\textit{s} bands.
The region between E$_\text{loss}$=0.5-1.2 eV corresponds to the optically forbidden crystal field \textit{dd} excitations, as widely reported in superconducting cuprates \cite{More11}, and identified in the DFT calculations below 1 eV. The orbital assignment of these excitations was verified by comparing their energy with the DFT calculations. 
Since the 3\textit{d} and 3\textit{p} bands of Cu and Cl atoms are disentangled in energy from the rest (Fig. \ref{Fig2}(b)), an expansion of the Bloch manifold can be performed in real space in terms of \textit{localized Wannier functions} \cite{Mar12}. A symmetry analysis of those Wannier functions allows to identify how the crystal field splitting affects the Cu 3\textit{d} orbitals. 

In a tetrahedral coordination, the 3\textit{d}$^9$ electronic configuration split the crystal field generated by Cl ions surrounding a Cu$^{2+}$ ion into the energetically lower Cu \textit{e} doublet and higher Cu \textit{t}$_{2}$ triplet. Owing to the Jahn-Teller uniaxial distortion of the tetrahedron, the \textit{t}$_{2}$ triplet is further split into the doubly degenerated \textit{d}$_{xz}$/\textit{d}$_{yz}$ states (\textit{e} irreducible representation) and the half-filled \textit{d}$_{xy}$ states (\textit{b}$_{2}$), and the degenerate \textit{d}$_{x^2-y^2}$ (\textit{b}$_{1}$) and \textit{d}$_{z^2}$ (\textit{a}$_{1}$) become also energetically non-equivalent (Fig. \ref{Fig1}(c)). The values of the crystal field splitting describing the tetrahedral distortion, \textit{D}$_q$, \textit{D}$_{\tau}$ and \textit{D}$_s$, are related to the on-site orbital energies through:


\begin{equation}\label{onsite_ener}
 \begin{split}
    a_{1}= 6\textit{D}_q-2\textit{D}_s-6\textit{D}_{\tau} \\
    b_{1}= 6\textit{D}_q+2\textit{D}_s-\textit{D}_{\tau} \\
    b_{2}= -4\textit{D}_q+2\textit{D}_s-\textit{D}_{\tau} \\
    e = -4\textit{D}_q-2\textit{D}_s+4\textit{D}_{\tau}
 \end{split}
\end{equation}
As shown in Fig. \ref{Fig1}(c), we found that the \textit{b}$_{1}$ and the \textit{b}$_{2}$ orbitals correspond to the lowest and highest energy levels, respectively. Apart from that, the strong orthorhombic distortion of the tetragonal environment inverts the energy levels of the doubly degenerated \textit{e} and \textit{a}$_{1}$.

To better understand the \textit{dd} excitations in the RIXS spectra, we adopt the \textit{hole} language, where the ground state represents a hole in the \textit{b}$_2$ orbital, hence, an \textit{e} orbital excitation corresponds to moving a hole from the \textit{b}$_2$ to the \textit{e} orbital. Therefore, the \textit{dd} transitions in Cs$_2$CuCl$_4$ originate from the decay of an electron from the \textit{b}$_1$, \textit{a}$_1$ and \textit{e} orbitals due to the broken degeneracy of the 3\textit{d} states. Within the energy resolution of our experimental setup, we can discriminate the 3 orbital intratomic transitions; 2 sharp excitations at 0.67 and 1.02 eV, respectively, and a shoulder at 1.21 eV (see Fig. \ref{Fig4}(b)). Having considered only the tetrahedral point group, \textit{T}$_d$, this would result in a crystal field splitting $\Delta_t$ of 1.2 eV, a rather high value assuming $\Delta_t\approx$0.44 $\Delta_{Oh}$. However, this number is a direct consequence of the strongly distorted tetrahedra towards a square planar geometry (see figure 1(b)), which increases the splitting of the energy levels, since $\Delta_{sp}\approx$1.74 $\Delta_{Oh}$. 
Further, we detect no orbital dispersion (Fig. 3(e)) indicating highly localized orbital excitations.
Since these \textit{dd} excitations are intra-atomic and well localized, they can be simulated within a full-multiplet calculation considering a single site of a Cu$^{2+}$ ions of the \textit{D}$_{4h}$ point group, isomorphic with the \textit{D}$_{2d}$ \cite{Ball62}, which is exemplified by a regular tetrahedron elongated along one of its \textit{C}$_2$ axes. The RIXS simulations were carried out with the exact diagonalization code Quanty \cite{Hav12,Lu14} including the Coulomb interactions and multiplet effects. We have adopted the \textit{U}$_{pd}$/\textit{U}$_{dd}$= 1.2 ratio for the Coulomb repulsion \cite{Zaan85,Hinkov20} and all atomic parameters were taken 80\% of the atomic Hartree-Fock values to compensate the absorption effects of the configuration interaction, leaving the radial integrals \textit{D}$_q$, \textit{D}$_s$ and \textit{D}$_{\tau}$, as input parameters in the calculation. The 3 orbital transitions corresponding to the \textit{b}$_1$, \textit{a}$_1$ and doubly degenerated \textit{e} orbitals are clearly identified in the calculated RIXS map of figure 4(a). As shown in Fig. 4(b), the experimental RIXS spectrum is fairly well reproduced with an additional instrumental and experimental broadening of 0.1 and 0.25 eV, respectively. 
From the simulation, we obtain \textit{D}$_q$=-0.120 eV, \textit{D}$_s$=0.085 eV and \textit{D}$_{\tau}$=-0.165 eV. The same sign of \textit{D}$_q$ and \textit{D}$_{\tau}$ indicates an axial elongation and, therefore, a proximity to a square planar geometry. The negative values of \textit{D}$_q$ and \textit{D}$_{\tau}$ account for the stabilization of the \textit{b}$_{1}$ state as the lowest level followed by the \textit{e} doublet, in agreement with the \textit{wannierization} of the DFT bands. Moreover, the value of \textit{D}$_s$ shows that the \textit{a}$_{1}$ level shifts its energy resulting in the splitting scheme depicted in figure 1(c). The ground state is 2-fold degenerate with spin and orbital quantum numbers $\langle\textit{S}_z\rangle=\pm 1/2$ and $\langle\textit{L}_z\rangle=\pm 0.31$ and orbital occupations N$_\text{Cu}$= 9.0 and N$_\text{Cl}$=6.0 electrons, which agree with the strong ionic character of the Cu$^{2+}$ and Cl$^-$ bond. Moreover, the Quanty simulation of the XAS spectrum using the same atomic Hartree-Fock and crystal field parameters is rather satisfactory (Figure 3(a)).

As shown in the Introduction in Fig. \ref{Fig1}, the half filled \textit{b}$_2$ orbital gives a straightforward explanation of the low temperature antiferromagnetism and magnetic anisotropy revealed by electron spin resonance (ESR) and neutron scattering \cite{Col03}. 
We can see in Fig. \ref{Fig1}e that the intrachain $J$ coupling has a direct overlap between the neighbouring \textit{b}$_2$ orbital along the \textit{b} axis. 
However, for the interchain neighbours defined by $J'$, we can see that the \textit{b}$_2$ orbitals of different chains are tilted (Fig. \ref{Fig1}f). 
This explains the observation of a stronger intrachain (along the \textit{b} axis) than interchain exchange coupling, \textit{J}'$\approx$\textit{J}/3.
The tilting of the \textit{b}$_2$ orbitals between different planes, as well as the larger separation between this kind of neighbours, agrees with the small exchange coupling observed for $J''$.

On the other hand, following theoretical predictions \cite{Brink04,Wohl13}, the observation of spin-orbital separation concentrated first on systems with orbital degeneracy \cite{Schla12,Ishi04} and revealed a sizable and asymmetric coupling between orbital and spin degrees of freedom in systems with  antiferromagnetic correlations. This coupling leads to the confinement of orbital excitations in magnetic systems with higher dimensions as reported in CaCu$_2$O$_3$ \cite{Bis15}, where magnetic correlations are still nearly one-dimensional. Therefore, the small exchange coupling energies and the directed character of the orbital motion along the chains would favor the dispersion of orbital excitations in Cs$_2$CuCl$_4$.
Besides, the local coupling of Jahn-Teller effect and lattice vibrations seems to be responsible for the absence of orbital waves in the 3D orbital ordered system KCuF$_3$ \cite{Li21}, despite of the Ising-like anisotropy in the short range order region \cite{Tow95}. Indeed, the Jahn-Teller effect is responsible for the high energy \textit{dd} orbital excitations but, to the best of our knowledge, no orbital ordering was observed in Cs$_2$CuCl$_4$. Therefore, relying on the directional character of electron hopping and the resulting one-dimensionality of orbital motion, the observation of collective orbital excitations in Cs$_2$CuCl$_4$ could be restricted to one orbital of particular symmetry and fall below the two-spinon continuum or limited by the instrumental resolution.

\section{Conclusions}

In conclusion, we have carried out a comprehensive spectroscopic study of the geometrically frustrated quantum spin liquid Cs$_2$CuCl$_4$, achieving a good agreement between experiments, cluster calculations and density functional theory. We have obtained the values of \textit{D}$_q$, \textit{D}$_s$ and \textit{D}$_{\tau}$ that fairly reproduce the RIXS spectrum and shown that the \textit{charge transfer gap}, commonly seen in the RIXS experiments, corresponds here to the energy gap between the CuCl$_4$ cluster and the Cs \textit{s} states. Moreover, the possible reason for the lack of an orbiton dispersion might be the strong ionic character of the Cu-Cl bond, unlike the orbiton dispersion observed in cuprates with a high degree of covalency \cite{Schla12,Bis15} or the low resolution of our instrumental setup.    

\section*{Acknowledgment}
We acknowledge Y. Lu and Justina Schlappa for fruitful discussions and critical reading of the manuscript.
S.B-C thanks the MINECO of Spain for financial support through the project PGC2018-101334-A-C22. This research was carried out at UE41 beamline at BESSY, a member of the Helmholtz Association (HGF). The research leading to this result has been supported by the project CALIPSOplus under the Grant Agreement 730872 from the EU Framework Programme for Research and Innovation HORIZON 2020. A.O.F. thanks MECD for the financial support received through the FPU grant FPU16/02572.




%

\end{document}